\documentclass{PoS}

\usepackage{amsmath}

\newcommand{\picpath}{.}

\title{QCD on the Cell Broadband Engine}

\ShortTitle{QCD on the Cell Broadband Engine}

\author{F.~Belletti$^a$,
  G.~Bilardi$^b$,
  M.~Drochner$^c$,
  N.~Eicker$^{d,e}$,
  Z.~Fodor$^{e,f}$,
  D.~Hierl$^g$,
  H.~Kaldass$^{h,i}$,
  T.~Lippert$^{d,e}$,
  T.~Maurer$^g$,
  \speaker{N.~Meyer}$^g$,
  A.~Nobile$^{j,k}$,
  D.~Pleiter$^i$,
  A.~Sch\"afer$^g$,
  F.~Schifano$^a$,
  H.~Simma$^{i,k}$,
  S.~Solbrig$^g$,
  T.~Streuer$^l$,
  R.~Tripiccione$^a$,
  T.~Wettig$^g$\\
  Email: \email{nils.meyer@physik.uni-regensburg.de}\\
  $^a$Department of Physics, University of Ferrara, 44100 Ferrara,
      Italy\\
  $^b$Department of Information Engineering,
      University of Padova, 35131 Padova, Italy\\ 
  $^c$ZEL, Research Center J\"ulich, 52425 J\"ulich, Germany\\
  $^d$ZAM, Research Center J\"ulich, 52425 J\"ulich, Germany\\
  $^e$Department of Physics, University of Wuppertal,
      42119 Wuppertal, Germany\\ 
  $^f$Institute for Theoretical Physics, Eotvos University, Budapest,
      Pazmany 1, H-1117, Hungary\\
  $^g$Department of Physics, University of Regensburg,
      93040 Regensburg, Germany\\ 
  $^h$Arab Academy of Science and Technology, P.O. Box 2033, Cairo, Egypt\\
  $^i$Deutsches Elektronen-Synchrotron DESY, 15738 Zeuthen, Germany\\
  $^j$European Centre for Theoretical Studies ECT$^\ast$, 13050
      Villazzano, Italy\\ 
  $^k$Department of Physics, University of Milano - Bicocca,
		20126 Milano, Italy\\  
  $^l$Department of Physics and Astronomy, University of Kentucky,
		Lexington, KY 40506-0055, USA
}

\abstract{We evaluate IBM's Enhanced Cell Broadband Engine (BE) as a
  possible building block of a new generation of lattice QCD machines.
  The Enhanced Cell BE will provide full support of double-precision
  floating-point arithmetics, including IEEE-compliant rounding. We
  have developed a performance model and applied it to relevant
  lattice QCD kernels. The performance estimates are supported by
  micro- and application-benchmarks that have been obtained on
  currently available Cell BE-based computers, such as IBM QS20 blades
  and PlayStation 3. The results are encouraging and show that this
  processor is an interesting option for lattice QCD applications.
  For a massively parallel machine on the basis of the Cell BE, an
  application-optimized network needs to be developed.}

\FullConference{The XXV International Symposium on Lattice Field Theory\\
  July 30 - August 4 2007\\
  Regensburg, Germany}

\begin{document}

\section{Introduction}

The initial target platform of the Cell BE was the PlayStation 3, but
the processor is currently also under investigation for scientific
purposes \cite{Williams,Nakamura}.  It delivers extremely high
floating-point (FP) performance, memory and I/O bandwidths at an
outstanding price-performance ratio and low power consumption.

We have investigated the Cell BE as a potential compute node of a
next-generation lattice QCD machine. Although the double precision
(DP) performance of the current version of the Cell BE is rather poor,
the announced Enhanced Cell BE version (2008) will have a DP
performance of $\sim 100$ GFlop/s and also implement IEEE-compliant
rounding. We have developed a performance model of a relevant lattice
QCD kernel on the Enhanced Cell BE and investigated several possible
data layouts. The applicability of our model is supported by a variety
of benchmarks performed on commercially available platforms.  We also
discuss requirements for a network coprocessor that would enable
scalable parallel computing using the Cell BE.

\section{The Cell Broadband Engine}

\begin{figure}[-b]
  \begin{center}
    \includegraphics[width=.85\textwidth]{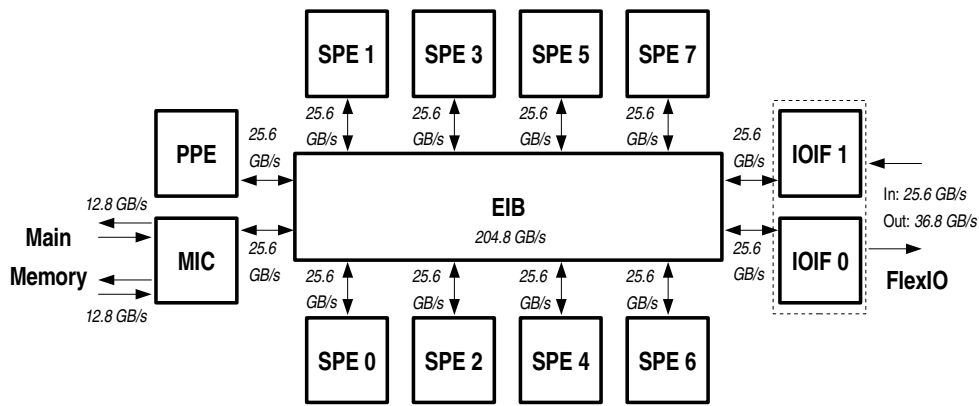}
    \caption{Main functional units of the Cell BE (see
      Ref.~\cite{Cell} for details).  Bandwidth values are given for a
      3.2 GHz system clock.}
    \label{fig:cell}
  \end{center}
\end{figure}

An introduction to the processor can be found in Ref.~\cite{Hofstee},
and a schematic diagram is shown in Fig.~\ref{fig:cell}. The
architecture is described in detail in Ref.~\cite{Cell}, and we only
give a brief overview here.

The Cell BE comprises one PowerPC Processor Element (PPE) and 8
Synergistic Processor Elements (SPE). In the following we will assume
that performance-critical kernels are executed on the SPEs and that
the PPE will execute control threads. Therefore, we only consider the
performance of the SPEs.  Each of the dual-issue, in-order SPEs runs a
single thread and has a dedicated 256 kB on-chip memory (local store =
LS) which is accessible by direct memory access (DMA) or by local
load/store operations to/from the 128 general purpose 128-bit
registers.  An SPE can execute two instructions per cycle, performing
up to 8 single precision (SP) operations. Thus, the aggregate SP peak
performance of all 8 SPEs on a single Cell BE is 204.8 GFlop/s at 3.2
GHz.\footnote{Available systems use clock frequencies of 2.8 or 3.2
  GHz. In our estimates we assume 3.2 GHz.}

The current version of the Cell BE has an on-chip memory controller
supporting dual-channel access to the Rambus XDR main memory (MM),
which will be replaced by DDR2 for the Enhanced Cell BE. The
configurable I/O interface supports a coherent as well as a
non-coherent protocol on the Rambus FlexIO channels.\footnote{In- and
  outbound bandwidths will be symmetric on the Enhanced Cell BE,
  namely 25.6 GB/s each.}  Internally, all units of the Cell BE are
connected to the coherent element interconnect bus (EIB) by DMA
controllers.

\section{Performance model}

To theoretically investigate the performance of the Cell BE, we use a
refined performance model along the lines of
Refs.~\cite{Bilardi,CellPerf}.  Our abstract model of the hardware
architecture considers two classes of devices: ($i$) \textit{Storage
  devices}: These store data and/or instructions (e.g., registers or
LS) and are characterized by their storage size.  ($ii$)
\textit{Processing devices}: These act on data (e.g., FP units) or
transfer data/instructions from one storage device to another (e.g.,
DMA controllers, buses, etc.) and are characterized by their
bandwidths $\beta_i$ and startup latencies $\lambda_i$.

An application algorithm, implemented on a specific machine, can be
broken down into different computational micro-tasks which are
performed by the processing devices of the machine model described
above.  The execution time $T_i$ of each task $i$\ is estimated by a
linear ansatz
\begin{equation}    	
  T_i \simeq I_i/\beta_i + \mathcal{O}(\lambda_i)\:,
  \label{eq:Timing}
\end{equation}
where $I_i$ quantifies the information exchange, i.e., the processed
data in bytes.

\begin{figure}
  \begin{center}
    \includegraphics[width=0.65\textwidth]{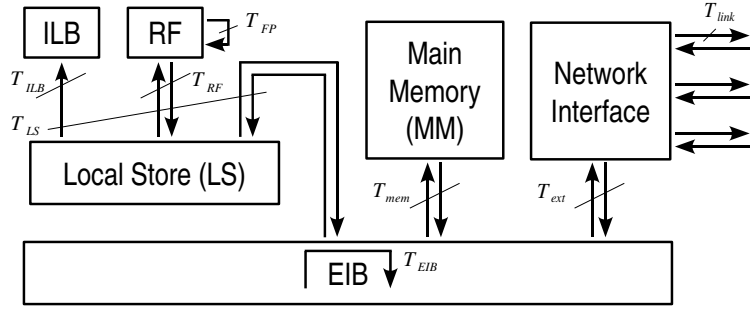}
    \caption{Data-flow paths and associated execution times $T_i$. For
      simplicity, only a single SPE is shown.}
    \label{fig:PerfModel}
  \end{center}
\end{figure}

Assuming that all tasks are running concurrently at maximal throughput
and that all dependencies (and latencies) are hidden by suitable
scheduling, the total execution time is
\begin{equation}    	
  T_\text{exe} \simeq \max_i T_i\:.
  \label{eq:Texe}
\end{equation}
We denote by $T_\text{peak}$ the minimal compute time for the FP
operations of an application that could be achieved with an ideal
implementation (i.e., saturating the peak FP throughput of the
machine, assuming also perfect matching between its instruction set
architecture and the computation).  The floating point efficiency
$\varepsilon_\text{FP}$ for a given application is then defined as
$\varepsilon_\text{FP} = T_\text{peak}/T_\text{exe}$.

In our analysis, we have estimated the execution times $T_i$ for data
processing and transport along all data paths indicated in
Fig.~\ref{fig:PerfModel}, in particular:
\begin{itemize}\itemsep-1mm
\item floating-point operations, $T_\text{FP}$
\item load/store operations between register file (RF) and LS, $T_\text{RF}$
\item off-chip memory access, $T_\text{mem}$
\item internal communications between SPEs on the same Cell BE, $T_\text{int}$
\item external communications between different Cell BEs, $T_\text{ext}$
\item transfers via the EIB (memory access, internal and external
  communications), $T_\text{EIB}$
\end{itemize}
Unless stated otherwise, all hardware parameters $\beta_i$ are taken
from the Cell BE manuals \cite{Cell}.

\section{Linear algebra kernels}

As a simple application of our performance model and to verify our
methodology, we analyzed various linear algebra computations. As an
example, we discuss here only a {\tt caxpy} operation: $c\cdot\psi +
\psi^\prime$ with complex $c$ and complex spin-color vectors $\psi$
and $\psi^\prime$.  If the vectors are stored in main memory (MM), the
memory bandwidth dominates the execution time, $T_\text{exe} \approx
T_\text{mem}$, and limits the FP performance of the {\tt caxpy} kernel
to $\varepsilon_\text{FP} \le 4.1$\%.  On the other hand, if the
vectors are held in the LS, arithmetic operations and LS access are
almost balanced ($T_\text{peak}/T_\text{LS} = 2/3$).  In this case, a
more precise estimate of $T_\text{FP}$ also takes into account
constraints from the instruction set architecture of the Cell BE for
complex arithmetics and yields a theoretical limit of
$\varepsilon_\text{FP} \le 50$\%.

We have verified the predictions of our theoretical model by
benchmarks on several hardware systems (Sony PlayStation 3, IBM QS20
Blade Server and Mercury Cell Accelerator Board).  In both cases (data
in MM and LS) the theoretical time estimates are well reproduced by
the measurements.  Careful optimization of arithmetic
operations\footnote{ We implemented our benchmarks of arithmetic
  operations in single precision. However, the theoretical analysis
  presented here refers to double precision on the Enhanced Cell BE.}
is required only in the case in which all data are kept in the LS (or,
in general, if $T_\text{exe} \approx T_\text{FP}$).
 
\section{Lattice QCD kernel}

The Wilson-Dirac operator is the kernel most relevant for the
performance of lattice QCD codes. We considered the computation of the
4-d hopping term
\begin{equation}
  \psi^\prime_x = \sum_{\mu=1}^4\left\{ U_{x,\mu}(1+\gamma_\mu)\psi_{x+\hat\mu} +
    U^\dagger_{x-\hat\mu,\mu}(1-\gamma_\mu)\psi_{x-\hat\mu}\right\}\:,
  \label{eq:Dirac}
\end{equation}
where $x = (x_1,x_2,x_3,x_4)$ is a 4-tuple of space-time coordinates
labeling the lattice sites, $\psi^\prime_x$ and $\psi_x$ are complex
spin-color vectors assigned to the lattice site $x$, and $U_{x,\mu}$
is an SU(3) color matrix assigned to the link from site $x$ in
direction $\hat\mu$.

The computation of Eq.~\eqref{eq:Dirac} on a single lattice site
amounts to 1320 floating-point operations.\footnote{We do not include
  sign flips and complex conjugation in the FLOP counting.}  On the
Enhanced Cell BE this yields $T_\text{peak} = 330$ cycles per site (in
DP). However, the implementation of Eq.~\eqref{eq:Dirac} requires at
least 840 multiply-add operations and $T_\text{FP} \ge 420$ cycles per
lattice site to execute. Thus, any implementation of
Eq.~\eqref{eq:Dirac} cannot exceed $78\%$ of the peak performance of
the Cell BE.

The time spent on possible remote communications and on load/store
operations for the op\-er\-ands ($9\times 12 + 8\times 9$ complex
numbers) of the hopping term \eqref{eq:Dirac} strongly depends on the
details of the lattice data layout. We assign to each Cell BE a local
lattice with $V_\text{Cell} = L_1\times L_2\times L_3\times L_4$
sites, and the 8 SPEs are logically arranged as $s_1\times s_2\times
s_3\times s_4 = 8$.  Thus, each single SPE holds a subvolume of
$V_\text{SPE} = (L_1/s_1)\times(L_2/s_2)\times(L_3/s_3)\times(L_4/s_4)
= V_\text{Cell}/8$ sites. Each SPE on average has $A_\text{int}$
neighboring sites on other SPEs {\em within} and $A_\text{ext}$
neighboring sites {\em outside} a Cell BE.

We consider a communication network with the topology of a 3-d torus.
We assume that the 6 inbound and the 6 outbound links can
simultaneously transfer data, each at a bandwidth of
$\beta_\text{link} = 1\;\text{GB/s}$, and that a bidirectional
bandwidth of $\beta_\text{ext} = 6\;\text{GB/s}$ is available between
each Cell BE and the network. This could be realized by attaching an
efficient network controller via the FlexIO interface.  We have
investigated different strategies for the lattice and data layout:
Either all data are kept in the on-chip local store of the SPEs, or
the data reside in off-chip main memory.

\subsection*{Data in on-chip memory (LS)}

We require that all data for a compute task can be kept in the LS of
the SPEs. Since loading of all data into the LS at startup is
time-consuming, the compute task should comprise a sizable fraction of
the application code. In QCD this can be achieved, e.g., by
implementing an entire iterative solver with repeated computation of
Eq.~\eqref{eq:Dirac}. Apart from data, the LS must also hold a minimal
program kernel, the run-time environment, and intermediate results.
Therefore, the storage requirements strongly constrain the local
lattice volumes $V_\text{SPE}$ and $V_\text{Cell}$.

The storage requirement of a spinor field $\psi_x$ is 24 real words
(192 Byte in double precision) per site, while a gauge field
$U_{x,\mu}$ needs 18 words (144 Byte) per link. Assuming that for a
solver we need storage corresponding to 8 spinors and $3\times 4$
links per site, the subvolume carried by a single SPE cannot be larger
than about $V_\text{SPE} = 79$ lattice sites.  Moreover, one lattice
dimension, say the 4-direction, must be distributed locally within the
same Cell BE across the SPEs (logically arranged as an $1^3\times8$
grid).  Then, $L_4$ corresponds to a global lattice extension and, as
a pessimistic assumption, may be as large as $L_4=64$. This yields a
very asymmetric local lattice\footnote{When distributed over 4096 Cell
  BEs, this corresponds to a global lattice size of $32^3\times64$.}
with $V_\text{Cell} = 2^3\times64$ and $V_\text{SPE} = 2^3\times8$.

\subsection*{Data in off-chip main memory (MM)}

When all data are stored in MM, there are no a-priori restrictions on
$V_\text{Cell}$.  On the other hand, we need to minimize redundant
memory accesses to reload the operands of Eq.~\eqref{eq:Dirac} into
the LS when sweeping through the lattice.  To also allow for
concurrent FP computation and data transfers (to/from MM or remote
SPEs), we consider a multiple buffering scheme.\footnote{In multiple
  buffering schemes several buffers are used in an alternating fashion
  to either process or load/store data. This requires additional
  storage (here in the LS) but allows for concurrent computation and
  data transfer.} A possible implementation of such a scheme is to
compute the hopping term \eqref{eq:Dirac} on a 3-d slice of the local
lattice and then move the slice along the 4-direction.  Each SPE
stores all sites along the 4-direction, and the SPEs are logically
arranged as a $2^3\times1$ grid to minimize internal and to balance
external communications between SPEs.  If the $U$- and $\psi$-fields
associated with all sites of {\em three} 3-d slices can be kept in the
LS at the same time, all operands in Eq.~\eqref{eq:Dirac} are
available in the LS. This optimization requirement again constrains
the local lattice size, now to $V_\text{Cell} \approx 800\times L_4$
sites.

\begin{table}[bt]
  \begin{center}
    \begin{tabular}{l|cc||cl|ccc}
      \multicolumn{2}{c}{data in on-chip LS} &
      \phantom{X} & \phantom{X} & 
      \multicolumn{4}{c}{data in off-chip MM}\\[2mm]\hline 
      $V_\text{Cell}$ & $2\times 2\times 2\times64$ &&& $L_1\times L_2\times L_3$ & 
      $8\times 8\times 8$ & $4\times 4\times 4$ & $2\times 2\times 2$\\\hline\hline 
      $A_\text{int}$  &  16       &&& $A_\text{int}/L_4$ & 48 & 12 & 3 \\
      $A_\text{ext}$  & 192       &&& $A_\text{ext}/L_4$ & 48 & 12 & 3 \\
      \hline
      $T_\text{peak}$ & 21        &&& $T_\text{peak}/L_4$ & 21 & 2.6 & 0.33\\
      \hline 
      $T_\text{FP}$   & 27        &&& $T_\text{FP}/L_4$ & 27 & 3.4 & 0.42\\
      $T_\text{RF}$   & 12        &&& $T_\text{RF}/L_4$ & 12 & 1.5 & 0.19\\
      $T_\text{mem}$  & ---       &&& $T_\text{mem}/L_4$ & 61 & 7.7 & 0.96\\
      $T_\text{int}$  &   2       &&& $T_\text{int}/L_4$ & 5 & 1.2 & 0.29\\
      $T_\text{ext}$  &  79       &&& $T_\text{ext}/L_4$ & 20 & 4.9 & 1.23\\
      $T_\text{EIB}$  &  20       &&& $T_\text{EIB}/L_4$ & 40 & 6.1 & 1.06\\
      \hline
      $\varepsilon_\text{FP}$   & \textbf{27\%}  &&& $\varepsilon_\text{FP}$ &
      \textbf{34\%}   & \textbf{34\%} & \textbf{27\%} 
    \end{tabular}
    \caption{Comparison of the theoretical time estimates $T_i$ (in
      1000 SPE cycles) for some micro-tasks arising in the computation
      of Eq.~\protect\eqref{eq:Dirac} for different lattice data
      layouts: keeping data either in the on-chip LS (left part) or in
      the off-chip MM (right part).  The first rows indicate the
      corresponding number of neighbor sites $A_\text{int}$ and
      $A_\text{ext}$.  Estimated efficiencies, $\varepsilon_\text{FP}
      = T_\text{peak}/\max_i T_i$, are shown in the last row.
      \vspace*{-1mm}}
   \label{tab:Timing}
  \end{center}
\end{table}

The predicted execution times for some of the micro-tasks considered
in our model are given in Table~\ref{tab:Timing} for both data layouts
and for reasonable choices of the local lattice size. If all data are
kept in the LS, the theoretical efficiency of about $27\%$ is limited
by the communication bandwidth ($T_\text{exe} \approx T_\text{ext}$).
This is also the limiting factor for the smallest local lattice with
data kept in MM, while for larger local lattices the memory bandwidth
becomes the limiting factor ($T_\text{exe} \approx T_\text{mem}$).

We have performed hardware benchmarks with the same memory access
pattern as \eqref{eq:Dirac}, using the above multiple buffering scheme
for data from MM.  We found that the execution times were at most 20\%
higher than the theoretical predictions for $T_\text{mem}$.

\section{Performance model and benchmarks for DMA transfers}

\begin{figure}
  \begin{center}
   \includegraphics[width=0.85\textwidth]{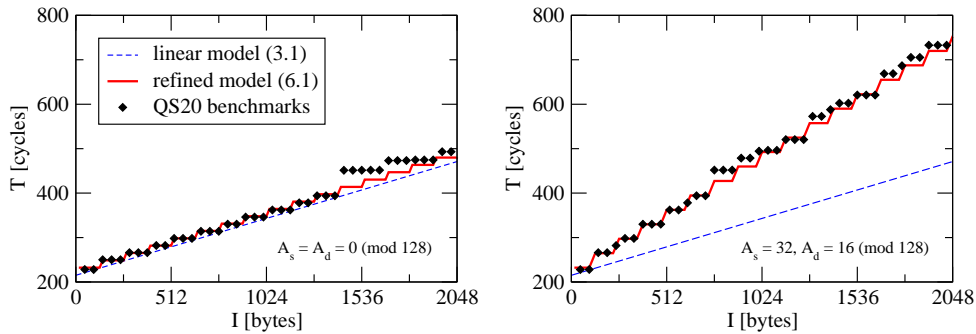}
   \caption{Execution time of LS-to-LS copy operations as a function
     of the transfer size.  In the left panel source and destination
     addresses are aligned, while in the right panel they are
     misaligned.  Filled diamonds show the measured values on an IBM
     QS20 system.  Dashed and full lines correspond to the theoretical
     prediction from Eq.~(\protect\ref{eq:Timing}) and
     Eq.~(\protect\ref{eq:LS_refined}), respectively.\vspace*{-2mm}}
  \label{fig:LsLsBench}
  \end{center}
\end{figure}
DMA transfers determine $T_\text{mem}$, $T_\text{int}$, and
$T_\text{ext}$, and their optimization is crucial to exploit the Cell
BE performance. Our analysis of detailed micro-benchmarks, e.g., for
LS-to-LS transfers, shows that the linear model Eq.~\eqref{eq:Timing}
does not accurately describe the execution time of DMA operations with
arbitrary size $I$ and address alignment.  We refined our model to
take into account the fragmentation of data transfers, as well as
source and destination addresses, $A_s$ and $A_d$, of the buffers:
\begin{equation}
  T_\text{DMA}(I,A_s,A_d) = \lambda^0 +
  \lambda^a \cdot N_a(I,A_s,A_d) +
  N_b(I,A_s) \cdot \frac{128\;\text{bytes}}{\beta}\:.
  \label{eq:LS_refined}
\end{equation}
Each LS-to-LS DMA transfer has a latency of $\lambda^0\approx
200\;\text{cycles}$ (from startup and wait for completion). The DMA
controllers fragment all transfers into $N_b$ 128-byte blocks aligned
at LS lines (and corresponding to single EIB transactions).  When
$\delta A = A_s-A_d$ is a multiple of 128, the source LS lines can be
directly mapped onto the destination LS lines. Then, we have $N_a=0$,
and the effective bandwidth $\beta_\text{eff} =
I/(T_\text{DMA}-\lambda^0)$ is approximately the peak value.
Otherwise, if the alignments do not match ($\delta A$ not a multiple
of 128), an additional latency of $\lambda^a \approx
16\;\text{cycles}$ is introduced for {\em each} transferred 128-byte
block, reducing $\beta_\text{eff}$ by about a factor of two.
Fig.~\ref{fig:LsLsBench} illustrates how clearly these effects are
observed in our benchmarks and how accurately they are described by
Eq.~\eqref{eq:LS_refined}.

\section{Conclusion and outlook}

Our performance model and hardware benchmarks indicate that the
Enhanced Cell BE is a promising option for lattice QCD.  We expect
that a sustained performance above $20\%$ can be obtained on large
machines.  A refined theoretical analysis, e.g., taking into account
latencies, and benchmarks with complete application codes are
desirable to confirm our estimate.  Strategies to optimize codes and
data layout can be studied rather easily, but require some effort to
implement.

Since currently there is no suitable southbridge for the Cell BE to
enable scalable parallel computing, we plan to develop a network
coprocessor that allows us to connect Cell BE nodes in a 3-d torus
with nearest-neighbor links.  This network coprocessor should provide
a bidirectional bandwidth of 1 GB/s per link for a total bidirectional
network bandwidth of 6 GB/s and perform remote LS-to-LS copy
operations with a latency of order 1 $\mu$s.  Pending funding
approval, this development will be pursued in collaboration with the
IBM Development Lab in B\"oblingen, Germany.

\end{document}